\documentclass[final,5p,times,twocolumn]{elsarticle}

\usepackage{graphicx,times}
\journal{Physica B}

\begin{document}  

\begin{frontmatter}

\title{Pseudo-zero-mode Landau levels and pseudospin waves in bilayer graphene}


\author{K. Shizuya}

\address{Yukawa Institute for Theoretical Physics,
Kyoto University,~Kyoto 606-8502,~Japan}

\begin{abstract}
Bilayer graphene in a magnetic field supports eight zero-energy Landau levels, 
which, as a tunable band gap develops, 
evolve into two nearly-degenerate quartets separated by the band gap.  
A close look is made into the properties of such an  isolated quartet 
of pseudo-zero-mode levels at half filling 
in the presence of an in-plane electric field and the Coulomb interaction. 
The pseudo-zero-mode levels turn out to support, via orbital mixing, charge carriers
with induced electric dipole moment, which lead to characteristic collective excitations, 
pseudospin waves, with some controllable features; possible experimental signals are discussed.
\end{abstract}


\end{frontmatter}

\section{Introduction}  
Graphene attracts a great deal of attention, both experimentally and
theoretically~\cite{NG,ZA},
for its exotic electronic transport.
Of particular interest is bilayer graphene~\cite{MF}, which has a unique property
that the (conduction-valence) band gap is controllable~\cite{OBSHR}
by use of external gates or chemical doping.

A particle-hole symmetric tower of Landau levels 
and the presence of some zero-energy levels
are features specific to graphene in a magnetic field.
Bilayer graphene  supports eight such zero-energy levels, 
which, as a tunable band gap develops, 
evolve into two nearly-degenerate quartets separated by the band gap.

In this paper we would like to report some unusual properties
of such an isolated quartet of pseudo-zero-mode levels,
especially, coherence and collective excitations  
in the presence of an in-plane electric field ${\bf E}_{\parallel}$ and the Coulomb
interaction~\cite{MS}.
We focus on revealing further controllable features in bilayer graphene
and point out the following:

(1) The pseudo-zero-mode levels, especially at half filling, 
support, via orbital level mixing, 
charge carriers with a nonzero electric dipole moment.
As a result, fine splitting of the pseudo-zero-mode quartet
 is also externally controlled by an in-plane electric field 
or by an injected Hall current.

(2) The interplay of the Coulomb interaction and an external field leads to rich spectra
of collective excitations, (orbital) pseudospin waves.

\section{Pseudo-zero-mode levels and dynamics}

Monolayer graphene supports as charge carriers massless Dirac fermions 
with a linear dispersion.
In bilayer graphene interlayer coupling modifies the intralayer "relativistic" spectra
to yield, in the low-energy branches, massive quasiparticles\cite{MF},
which, in a magnetic field $B$, lead to a tower of Landau levels 
$|n,y_{0}\rangle$ of energy 
\begin{eqnarray}
\epsilon_{n} =  {\rm sgn}[n]\, \omega_{\rm c}^{\rm bi} \sqrt{|n| (|n| -1) } ,  
\label{specnonzero}
\end{eqnarray}
labeled by integers $n=\pm 0, \pm 1,  \pm2, \dots$, 
the center coordinate  $y_{0} \equiv \ell^{2} p_{x}$
with the magnetic length $\ell= 1/\sqrt{eB}$
and the characteristic cyclotron energy 
$\omega^{\rm bi}_{c}\sim  4\times  B[{\rm T}] $ meV   (with  $B$ in tesla).

The $|n|=0$ and $|n|=1$ levels have zero energy,
and, with  a band gap $U>0$, evolve into
two sectors of nearly-degenerate pseudo-zero-mode levels
(separated by a band gap and in valley),
with spectra
\begin{eqnarray}
\epsilon_{n=\pm0} &=&  \pm \textstyle{1\over{2}}\, U,\  
\epsilon_{n=\pm1} =  \pm ({1\over{2}}\, U -2m ).    
\label{zeromodespectrum}
\end{eqnarray}
where $m\approx 0.006\times B[{\rm T}]\, U \ll U$ denotes a fine splitting 
within each sector.


\begin{figure}[tbp]
\begin{center}
\includegraphics[scale=.3]{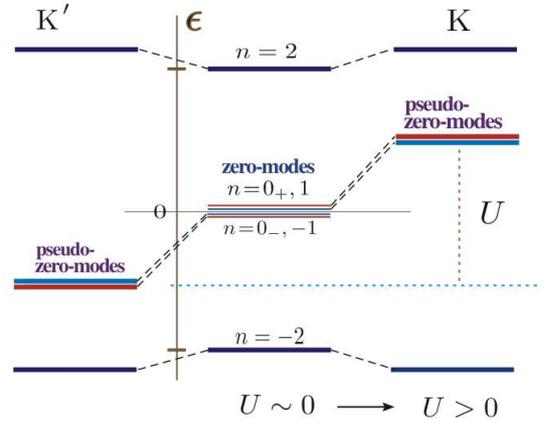}
\end{center}
\vskip-0.5cm
\caption{ Pseudo-zero-mode Landau levels }
\end{figure}


The presence of the zero-energy levels
and their double-fold degeneracy (per spin and valley)  have
a topological origin.
This degeneracy is affected by a band gap $U$ and 
an external field $A_{0}$.
Accordingly, the pseudo-zero-mode levels have  
an intrinsic tendency to be degenerate, but its fine structure depends sensitively 
on the environment.

The purpose of the present paper is to study 
such controllable features of the pseudo-zero-mode quartet 
in the presence of external fields and Coulomb interactions. 
We are particularly interested 
in the properties of this quartet at half filling,
where mixing of the pseudo-zero-modes may take place.
For definiteness we focus on the $n= (0_{+}, 1)$ sector at one valley,
i.e., around filling factor $\nu=2$ (or $\nu=1$ when the spin is resolved),
and ignore the presence of other levels 
which are separated by relatively large gaps.

The pseudo-zero-mode levels of our interest contain  
two different Landau orbitals $n=0_{+}$ and $n=1$ (per spin).
Let us put these modes,  
$\psi_{0_{+}}(y_{0},t)$ and $\psi_{1}(y_{0},t)$,
into a two-component spinor $\Psi=(\psi_{0_{+}}, \psi_{1} )^{\rm t}$,
and define the pseudospin operators in the $(0_{+}, 1)$ orbital space,
\begin{eqnarray}
S^{\mu}_{\bf -p} &=&\gamma_{\bf p} \int dy_{0}\, 
\Psi^{\dag}\, {1\over{2}}\,\sigma^{\mu} e^{i{\bf p \cdot r}}\Psi,
\label{Smup}
\end{eqnarray}
with Pauli matrices $\sigma^{a}$ and $\sigma^{0}=1$;
$\gamma_{\bf p}= e^{- \ell^{2}{\bf p}^{2}/4}$.
The charge density 
$\rho_{-{\bf p}}=\int d^{2}{\bf x}\,  e^{i {\bf p\cdot x}}\,\psi^{\dag}\psi$ 
 projected to the $(0_{+}, 1)$ sector 
is thereby written as
\begin{equation}
\bar{\rho}_{\bf -p} 
= 2\, ( w^{0}_{\bf p}S^{0}_{\bf -p} + w^{3}_{\bf p} S^{3}_{\bf -p}
 + w^{1}_{\bf p} S^{1}_{\bf -p}+ w^{2}_{\bf p}S^{2}_{\bf -p} ), 
\end{equation}
where $w^{0}_{\bf p} = 1- \ell^{2}{\bf p}^{2}/4$,
 $w^{3}_{\bf p} = \ell^{2}{\bf p}^{2}/4$,
 $w^{1}_{\bf p} = i\ell p_{y}/\sqrt{2}$, and 
$w^{2}({\bf p}) = i\,\ell p_{x}/\sqrt{2}$.

The relevant part of the bilayer  Hamiltonian 
governing the electrons in the $(0_{+}, 1)$ sector is written as
\begin{eqnarray}
\triangle \bar{H}=-\int\! d^{2}{\bf x}\,  e A_{0}\, \bar{\rho} + 
\int\! dy_{0}\, m\, \Psi^{\dag}\sigma^{3}\, \Psi,
\end{eqnarray}
where $A_{0}$ stands for an external potential 
and $m \approx 0.006\times B[{\rm T}]\, U$ for the fine level 
splitting.
One can equally write
\begin{eqnarray}
\triangle \bar{H} 
&=& 2\sum_{\bf p} \, {\cal P}^{\mu}_{\bf p}\, S^{\mu}_{\bf -p},\\
({\cal P}^{0}, {\cal P}^{3})\! \!&\approx&\! \!(- m -eA_{0}, m) , \ 
({\cal P}^{2},  {\cal P}^{1}) = {e\ell\over{\sqrt2}}\,( E_{x}, E_{y}),  \ \
\end{eqnarray}
where ${\bf E}_{\parallel} =(E_{x}, E_{y}) = -\partial_{\bf x}A_{0}$ 
denotes the in-plane electric field.
For the Coulomb interaction one can retain only the intralayer
interaction since the pseudo-zero-modes at each valley essentially   
lie on the same layer,
\begin{equation}
\bar{H}^{\rm C} 
= {1\over{2}} \sum_{\bf p}
v_{\bf p} :\bar{\rho}_{\bf -p}\, \bar{\rho}_{\bf p}: ,
\label{Hcoul}
\end{equation}
where 
$v_{\bf p}= 2\pi \alpha/(\epsilon_{\rm b} |{\bf p}|)$ 
with 
$\alpha = e^{2}/(4 \pi \epsilon_{0}) \approx 1/137$ and 
the substrate dielectric constant $\epsilon_{\rm b}$.

\section{Pseudospin textures}

In this section we study the properties of the pseudo-zero-mode levels
at half filling, using the projected Hamiltonian  
$\bar{H} \equiv \triangle \bar{H} + \bar{H}^{\rm C}$, 
with focus on orbital mixing of the zero-modes.

Let us suppose that the half-filled state $|G\rangle$
is given by a configuration where the pseudospin points in  
a fixed direction in pseudospin space, i.e., 
$\langle G|S^{a}_{\bf p=0}|G\rangle = {1\over{2}}\,
N_{e}\, n^{a}$ and $n^{a}n^{a} =1$ with the total number 
of electrons $N_{e}= 2\,\langle G| S^{0}_{\bf p=0}|G\rangle$.
For $n^{a}$ we use the parametrization 
$n^{1}= \sin\theta\, \cos \phi$, $n^{2}= \sin\theta\, \sin \phi$, 
$n^{3}= \cos \theta$, with $-\pi < \theta \le \pi$ and $0\le \phi \le  \pi$.

Note that $n^{3}=1$ corresponds to the filled $n=0_{+}$ level 
while $n^{3} = -1$ represents the filled $n=1$ level.
The direction ${\bf n}=(n^{1},n^{2},n^{3})$ would, in general, 
vary in response to the external field $A_{0}$, and,
as ${\bf n}$ tilts from $n^{3}=\pm 1$, 
the $n=0_{+}$ and $n=1$ levels start to mix.
For selfconsistency we assume that  $A_{0}$ represents a uniform in-plane electric 
field ${\bf E}_{\parallel} = -\partial_{\bf x}A_{0}$ and that it leads to 
a homogeneous state $|G\rangle$ of uniform density $\rho_{0}= \nu/(2\pi \ell^{2})$
(with $\nu=2$ for the spin-degenerate $\nu=2$ state).
We thus consider all configurations with $\langle G|S^{a}_{\bf p=0}|G \rangle = {1\over{2}}\,
N_{e}\, n^{a}$ and single out the ground state or the associated  $n^{a}$ 
by minimizing the energy $\langle G| \bar{H} |G \rangle$.

The calculation of the Coulomb energy 
$\langle G| \bar{H}^{\rm C} |G\rangle$,
in particular, requires the knowledge of pseudospin structure factors 
$\langle G|S^{\mu}_{\bf p} S^{\nu}_{\bf q}|G \rangle$, which, 
for the present half-filled state with pseudospin $\propto n^{a}$, 
are given in Ref.~[5].
The energy  is conveniently written in ${\bf x}$ space as
$\langle G| \triangle\bar{H} + \bar{H}^{\rm C} |G\rangle
=  \rho_{0}\int d^{2}{\bf x}\, {\cal H}_{\rm eff}$ with
\begin{eqnarray}
 {\cal H}_{\rm eff}
&=& - e A_{0} + E(\theta) - \textstyle {1\over{2}}\, V_{1},   \nonumber\\ 
E(\theta)
&=& 
 m\, (\cos \theta -1) + {\cal E}\,  \sin\theta 
+ \textstyle{1\over{32}}\,V_{1}\, (1- \cos \theta)^{2},
\label{Etheta}
\end{eqnarray}
where ${\cal E} = (e\ell/\sqrt{2})\, 
{\bf E}_{\parallel}\cdot \hat{\bf n}_{\parallel}$ and 
$\hat{\bf n}_{\parallel}=(\hat{n}_{x}, \hat{n}_{y}) 
= (\sin \phi, \cos \phi)$
is an in-plane unit vector;
\begin{equation}
V_{1} = \sqrt{\pi/2}\, \alpha/(\epsilon_{b}\, \ell)
\end{equation}
sets the scale of the Coulomb exchange energy.
In the present notation 
${\cal H}_{\rm eff}$ stands for energy per electron 
in state $|G\rangle$.

 Note first that the half-filled pseudo-zero-mode state 
has an in-plane electric dipole moment
\begin{eqnarray}
{\bf d}_{e}&=& -{e\ell\over{\sqrt{2}}}\, (n^{2}, n^{1}) 
= -{e\ell\over{\sqrt{2}}}\, \sin\theta\,  \hat{\bf n}_{\parallel}
\end{eqnarray}
of strength $(e\ell/\sqrt{2})\, |\sin\theta|$ per electron, 
proportional to the in-plane component $(n^{2}, n^{1})$ of the pseudospin.
Mixing of the $n=0_{+}$ and $n=1$ modes gives rise to this dipole.  
Note also that the Coulomb correlation energy $\propto V_{1}$ is symmetric about 
the $n^3$ axis $\parallel {\bf B}$.  It alone favors $\theta=0$, 
the filled $n=0_{+}$ level, and varies continuously 
by an amount $\triangle E_{c}= (1/8)\, V_{1}$ 
as ${\bf n}$ sweeps in pseudospin space.
The Coulomb interaction thus significantly enhances 
the splitting of the pseudo-zero-mode levels.

Obviously the energy is lower when when
${\bf n}_{\parallel} \parallel {\bf E}_{\parallel}$,
or ${\cal E} =  e\ell |{\bf E}_{\parallel}|/\sqrt{2}$.
Accordingly it is    convenient, without loss of generality, to suppose that 
the in-plane field ${\bf E}_{\parallel}$ and ${\bf n}_{\parallel}$ 
lie along the $y$ axis,
${\cal E} = e \ell E_{y}/\sqrt{2} \ge 0$.
With this choice the "1", "2" and "3" axes in pseudospin space coincide 
with the $y$, $x$ and $-z$ axes in real space, respectively.
We henceforth adopt this choice and set 
$(n^{1}, n^{2}, n^{3}) = ( \sin \theta, 0,\cos \theta)$.

Let us explore stable configurations of the half-filled zero-mode state
in the next section and here study collective excitations 
over a given ground state $|G\rangle|_{\bf n}$.
We focus on a special class of low-energy collective excitations,
pseudospin waves, that are rotations 
about the energy minimum $|G\rangle|_{\bf n}$ in pseudospin space.
Such a collective state is represented as a texture state
\begin{eqnarray}
|\tilde{G}\rangle = e^{-i{\cal O}} |G\rangle|_{\bf n},
\end{eqnarray}
where the operator 
${\cal O}= \sum_{\bf p}\gamma_{\bf p}^{-1}\,
\Omega^{a}_{\bf p}\, S^{a}_{\bf -p}$ 
locally tilts the pseudospin from ${\bf n}$ 
by small angle $\Omega_{\bf p}$.
The energy $\langle \tilde{G}|\bar{H}|\tilde{G}\rangle
=\langle G|e^{i{\cal O}}\bar{H} e^{-i{\cal O}}|G\rangle$
of the excited state is a functional of $\Omega^{a}_{\bf p}$,
and serves as an effective Hamiltonian.
The associated Lagrangian is neatly written as
\begin{equation}
\rho_{0}\!\!\int\! d^{2}{\bf x}\, L_{\rm eff}=\langle G|e^{i{\cal O}}\, 
(i \partial_{t} -\bar{H}) e^{-i{\cal O}}|G\rangle.
\end{equation}
Actual calculation is made in a systematic way if one notes that  
the pseudospin operators $S^{\mu}_{\bf p}$ in Eq.~(\ref{Smup}) obey 
the SU(2)$\times W_{\infty}$ algebra.
The result is an effective Lagrangian of the form~\cite{MS}
\begin{eqnarray}
L_{\Phi} &=& {1\over{2}}\, (\partial_{t} \Phi)^{2} 
- {1\over{2}}\, \Phi\,  (M_{\bf p})^{2}\, \Phi,
\nonumber\\ 
M_{\bf p}
&=& 2\, \sqrt{(\kappa_{\eta}^{2}+ F_{\bf p})(\kappa_{\zeta}^{2}+ G_{\bf p})
 - |W_{\bf p}|^{2}},\nonumber\\ 
\kappa_{\zeta}^{2} &=&  E''(\theta),\ \
\kappa_{\eta}^{2} = -\, {\cal E}/\sin\theta, 
\label{LMp}
\end{eqnarray}
where $(F_{\bf p}$, $G_{\bf p}$, $W_{\bf p}) \sim O(V_{1})$ 
go to zero for ${\bf p}\rightarrow 0$.  
 The spectrum $M_{\bf p}$ of pseudospin waves is 
in general anisotropic in ${\bf p}$ and depends 
critically on the stable-state configuration ${\bf n}$.
In particular, $M_{{\bf p}\rightarrow 0}= 2\kappa_{\zeta}\kappa_{\eta}$
and $M_{{\bf p}\rightarrow \infty} 
\approx V_{1}/2 + \kappa_{\zeta}^{2}+ \kappa_{\eta}^{2}$.
With $\theta\rightarrow 0$ and ${\cal E}\rightarrow 0$, our $M_{\bf p}$ precisely reproduces an excitation spectrum derived in Ref.~\cite{BNM} 
by assuming spatial isotropy.

Note here that the Coulomb interaction alone 
yields $\kappa_{\eta}^{2}=0$.
This implies that, unlike in ordinary bilayer quantum Hall systems, 
there is no cost of interlayer capacitance energy. 
Coherence is thus easier to form in 
the pseudo-zero-mode sector of bilayer graphene.

\section{ground states and pseudospin waves}

Let us now examine orbital configurations  of 
the half-filled pseudo-zero-mode state  
and the associated pseudospin waves. Note first that
the Coulomb correlation energy
\begin{eqnarray}
\triangle E_{c} = {1\over{8}}\, \sqrt{{\pi\over{2}}}\, 
{\alpha\over{\epsilon_{b}\,\epsilon_{\rm sc}\, \ell}}
\approx (2.2/ \epsilon_{\rm sc})\, \sqrt{B[{\rm T}]}\
{\rm meV}
\label{deltaEc}
\end{eqnarray}
with a typical value $\epsilon_{b} \sim 4$ 
sets the basic energy scale;
here $\epsilon_{\rm sc}$ effectively takes account of the effect of screening 
coming from vacuum (Dirac-sea) polarization~\cite{MS}, specific to graphene.
In ratio the intrinsic zero-mode level gap 
$2m$ 
and the in-plane field ${\cal E}= e \ell |{\bf E}_{\parallel}|/\sqrt{2}$ 
read 
\begin{eqnarray}
m/\triangle E_{c} &\approx& 
3\, \epsilon_{\rm sc}\times 10^{-3}\,  \sqrt{B[{\rm T}]}\, U[{\rm meV}],\\
{\cal E}/\triangle E_{c} &\approx& 0.9\, \epsilon_{\rm sc}
\times 10^{-3}\,   E[{\rm V/cm}]/B[{\rm T}].
\end{eqnarray}

One has to look for the minimum of $E(\theta)$, 
$E_{\rm min} = E(\theta_{\rm min})$,
to determine the orbital configuration of $|G\rangle$.
Let us begin with the case where ${\bf E}_{\parallel}$ is absent.
(i) For $m > \triangle E_{c}$ (though rather unrealistic), 
one finds $E_{\rm min}$ at $\theta = \pi$, i.e., 
the filled  $n=1$ level is realized.
The pseudospin waves have a finite energy gap
$M_{{\bf p}=0} = 2\, (m- \triangle E_{c}) > 0$
and the spectrum is isotropic.

On the other hand, (ii) for $0 < m < \triangle E_{c}$, one finds
$E_{\rm min}$ at $\theta =\pm\theta_{\rm min}$;
$\theta_{\rm min}$ varies from 0 to $\pi$ with increasing $m$. 
Here we encounter a somewhat strange situation: 
The rotational invariance (about the applied field  ${\bf B}$) is spontaneously broken 
and the pseudospin waves are gapless.

This actually signals an instability.
In this $m$ range the assumed ground state $|G\rangle$, taken to be 
homogeneous in space, acquires spontaneous
in-plane electric polarization $\propto \sin \theta_{\rm min}$.
Such electrically-polarized  {\sl homogeneous} configurations, 
unless polarization is relatively weak, are unstable 
against local charge inhomogeneities 
and would decay into {\sl inhomogeneous} configurations.
We thus speculate that the half-filled state 
in the realistic $0 < m < \triangle E_{c}$ range may 
form many domains about local charge excesses for stabilization.

(iii) The in-plane field ${\cal E}= e \ell E_{y}/\sqrt{2} >0$ tilts the pseudospin 
toward $\theta = -\pi/2$ and competes with $\triangle E_{c}$.
The charge carriers thereby acquire a nonzero electric dipole moment 
$\propto \sin \theta$
and the pseudospin waves always have 
a finite excitation gap.
The potential instability of the texture state, mentioned in (ii),  weakens 
and eventually disappears for large $|{\bf E}_{\parallel}|$.
In Fig.~2 we plot the excitation gap $M_{{\bf p}=0}$ as a function of 
$|{\bf E}_{\parallel}|$ for some typical values of 
$B[{\rm T}]$ and $m\propto U[{\rm meV}]$;
$M_{{\bf p}=0}$ falls in the frequency range of microwaves.


\begin{figure}[tbp]
\begin{center}
\includegraphics[scale=.95]{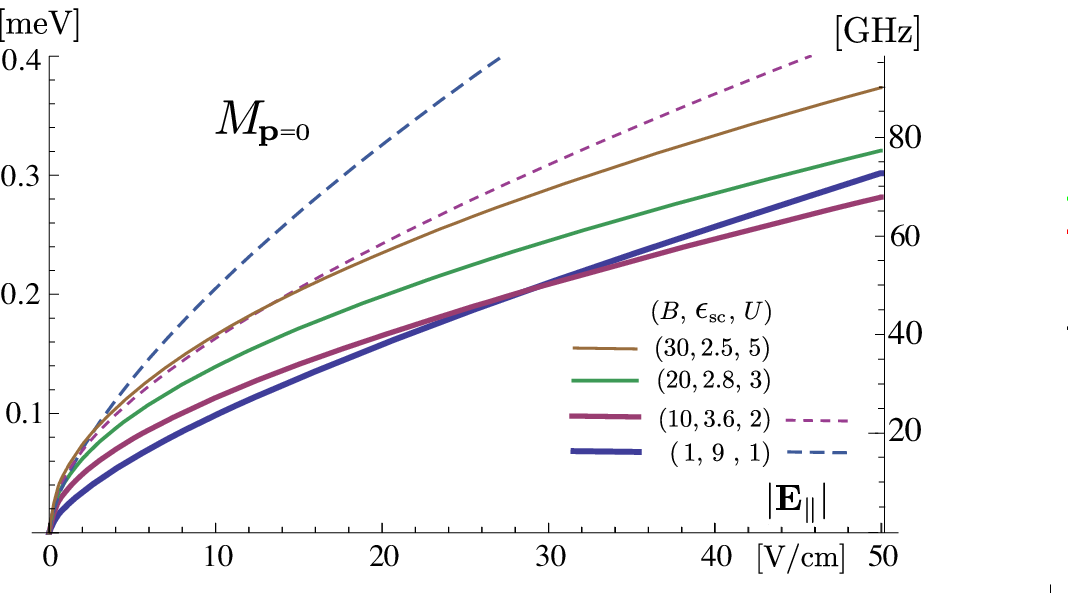}
\end{center}
\caption{
Pseudospin-wave excitation gap $M_{{\bf p}=0}$, in units of meV and GHz, plotted 
as a function of the in-plane field $|E_{\parallel}|$ 
for some typical values of magnetic field $B[{\rm T}]$ and band gap $U[{\rm meV}]$.
Dashed curves refer to the cases 
where the screening effect is turned off, 
$\epsilon_{\rm sc}\rightarrow 1$.
}
\end{figure}


If ${\bf E}_{\parallel}$ is sufficiently strong, 
a sizable gap $M_{{\bf p}=0}$
would arise, leading to an incompressible $\nu=2$ state.  
One could thereby observe the (spin-degenerate) $\nu=2$ Hall plateau 
with a suitably strong injected current. 
${\bf E}_{\parallel}$ is practically strong for ${\cal E}\gtrsim 0.3\, \triangle E_{c}$,
where $M_{{\bf p}=0} = 2\, \kappa_{\zeta}\kappa_{\eta} 
\approx 2 {\cal E} + {1\over{2}}\,  \triangle E_{c}$
with $\kappa_{\zeta}^{2} \approx  {\cal E}  +  {1\over{2}}\,  \triangle E_{c}$
and $\kappa_{\eta}^{2} \approx  {\cal E}$.

It  would be a challenge to directly detect the pseudospin-wave excitation gap
by microwave absorption or reflection experimentally. 
One may follow Ref.~[5] and derive the optical conductance due to transitions 
within the $(0_{+}, 1)$ sector,  
\begin{eqnarray}
\triangle\sigma_{xx}(\omega) &=&   (\nu\, e^{2}/\hbar)\,
\kappa_{\zeta}^2\,   \delta (\omega - M_{{\bf p}=0}), \nonumber \\
\triangle\sigma_{yy}(\omega) &=&  (\nu\, e^{2}/\hbar)\,
(\cos \theta_{\rm min})^2\, \kappa_{\eta}^2\, \delta (\omega - M_{{\bf p}=0}),
\end{eqnarray}
which, with disorder taken into account, is significantly peaked 
around  $\omega \sim  M_{{\bf p}=0}$.  Note that the optical conductance  
is spatially anisotropic, with the larger component $\triangle\sigma_{xx}(\omega)$
perpendicular to ${\bf E}_{\parallel}$ (or along an injected current).

\section{Summary and discussion}

Zero-mode Landau levels, specific to graphene in a magnetic field, 
deserve attention in their own right.
In bilayer graphene they evolve, with a tunable band gap,  
into two quartets of nearly-degenerate pseudo-zero-mode levels,
which, unlike in monolayer graphene, involve two different 
orbital indices $n=0, 1$. One would thus expect 
interesting quantum phenomena of orbital level mixing.

In this paper we have studied the effects of 
an external field and the Coulomb interaction 
on the pseudo-zero-mode quartet.
This quartet, especially at half filling, supports, via orbital mixing,
quasiparticles with charge and electric dipole,
which give rise to characteristic collective excitations, 
pseudospin waves.

The pseudospin-wave excitation gap $M_{{\bf p}=0}$ is generally small, 
reflecting the intrinsic degeneracy of the pseudo-zero-mode levels,
but turns out to depend sensitively on an in-plane field ${\bf E}_{\parallel}$.
This means that the gap  $M_{{\bf p}=0}$ is tunable by an in-plane field or 
by an injected current.

An experimental signature of the field-induced gap is 
to observe the quantum Hall effect 
with an injected current; one would be able 
to resolve the $\nu= \pm 2$ Hall plateaus 
(or the spin-resolved $\nu=\pm1$ plateaus)
using a suitably strong current.
A direct study of the excitation gap $M_{{\bf p}=0}$ and 
its field dependence by microwave absorption or reflection 
would also clarify the unique controllable features of 
the pseudo-zero-mode sector in bilayer graphene.  
\\

This work was supported in part by a Grant-in-Aid for Scientific Research
from the Ministry of Education, Science, Sports and Culture of Japan 
(Grant No. 21540265).


\end{document}